\documentclass[useAMS,usenatbib]{mn2e}

\usepackage{natbib}
\usepackage{graphicx}
\usepackage{color}
\usepackage{xspace}
\usepackage{xargs}
\usepackage{lineno}
\usepackage{footnote} 
\makesavenoteenv{tabular}

\usepackage[usenames,dvipsnames]{xcolor}

\newcommand{\B}[1]{{ {#1}}}
\newcommand{\myemail}{Rebecca.Kennedy@nottingham.ac.uk}
\newcommand{\galapagos}{{\sc galapagos}\xspace}
\newcommand{\galapagosII}{{\sc galapagos-2}\xspace}
\newcommand{\galfit}{{\sc galfit}\xspace}

\newcommand{\galfitm}{{\sc galfitm}\xspace}
\newcommand{\megamorph}{{MegaMorph}\xspace}
\newcommand{\sersic}{S\'ersic\xspace}

\newcommandx{\N}[2][1= ,2= ]{$\mathcal{N}^{#1}_{#2}$\xspace}
\newcommandx{\R}[2][1= ,2= ]{$\mathcal{R}^{#1}_{#2}$\xspace}
\newcommand{\re}{R_{\rm e}}

\title[Wavelength dependence of galaxy structure]{Galaxy And Mass Assembly (GAMA): the wavelength dependence of galaxy structure versus redshift and luminosity}
\author[Kennedy et al.]{Rebecca~Kennedy$^1$\thanks{E-mail: \myemail}, Steven~P.~Bamford$^1$, Ivan Baldry$^2$, Boris~H\"au\ss ler$^{3,4}$,\newauthor Benne~W.~Holwerda$^5$, Andrew~M.~Hopkins$^6$, Lee~S.~Kelvin$^7$, Rebecca~Lange$^8$, \newauthor Amanda~J.~Moffett$^8$, Cristina~C.~Popescu$^{9,10}$, Edward~N.~Taylor$^{11}$,\newauthor Richard~J.~Tuffs$^{12}$, Marina~Vika$^{13}$, Benedetta~Vulcani$^{14}$
\smallskip\\
$^{1}$School of Physics \& Astronomy, The University of Nottingham, University Park, Nottingham, NG7 2RD, UK\\
$^2$ARI, Liverpool John Moores University, IC2, Liverpool Science Park, 146 Brownlow Hill, Liverpool, L3 5RF\\
$^{3}$University of Oxford, Denys Wilkinson Building, Keble Road, Oxford, Oxon OX1 3RH, UK\\
$^{4}$University of Hertfordshire, Hatfield, Hertfordshire AL10 9AB, UK\\
$^{5}$University of Leiden, Sterrenwacht Leiden, Niels Bohrweg 2, NL-2333 CA Leiden, The Netherlands\\
$^{6}$Australian Astronomical Observatory, PO Box 915, North Ryde, NSW, NSW 1670, Australia\\
$^{7}$Institut f\"{u}r Astro- und Teilchenphysik, Universit\"{a}t Innsbruck, Technikerstra{\ss}e 25, 6020 Innsbruck, Austria\\
$^{8}$ICRAR, University of Western Australia, Crawley, WA 6009, Australia\\
$^{9}$Jeremiah Horrocks Institute, University of Central Lancashire, PR1 2HE, Preston, UK\\
$^{10}$The Astronomical Institute of the Romanian Academy, Str. Cutitul de Argint 5, Bucharest, Romania\\
$^{11}$School of Physics, the University of Melbourne, 3010 VIC, Australia\\
$^{12}$Max Planck Institute fuer Kernphysik, Saupfercheckweg 1, 69117 Heidelberg, Germany\\
$^{13}$IAASARS, National Observatory of Athens, GR-15236 Penteli, Greece\\
$^{14}$Kavli Institute for the Physics and Mathematics of the Universe (WPI), UTIAS, the University of Tokyo, Kashiwa, 277-8582, Japan\\
}
\begin{document}
\date{Accepted ... Received ...; in original form ...}
\maketitle

\begin{abstract}
We study how the sizes and radial profiles of galaxies vary with wavelength, by fitting \sersic functions simultaneously to imaging in nine optical and near-infrared bands.
To quantify the wavelength dependence of effective radius we use the ratio, \R, of measurements in two restframe bands.
The dependence of \sersic index on wavelength, \N, is computed correspondingly.
\citet{Vulcani2014} have demonstrated that different galaxy populations present sharply contrasting behaviour in terms of \R and \N.
Here we study the luminosity dependence of this result.  We find that at higher luminosities, early-type galaxies display a more substantial decrease in effective radius with wavelength, whereas late-types present a more pronounced increase in \sersic index. The structural contrast between types thus increases with luminosity.

By considering samples at different redshifts, we demonstrate that lower data quality reduces the apparent difference between the main galaxy populations.  However, our \B{conclusions remain} robust to this effect. 

\B{We show that accounting for different redshift and luminosity selections partly reconciles the size variation measured by \citeauthor{Vulcani2014} with the weaker trends found by other recent studies.}
Dividing galaxies by visual morphology confirms the behaviour inferred using morphological proxies, although the sample size is greatly reduced.

Finally, we demonstrate that varying dust opacity and disc inclination can account for features of the joint distribution of \R and \N for late-type galaxies. However, dust does not appear to explain the highest values of \R and \N. The bulge-disc nature of galaxies must also contribute to the wavelength-dependence of their structure.

\end{abstract}

\section{Introduction}

The assembly history of a galaxy is written in its stellar populations. Their colours (and spectral features) reflect the time and conditions of star formation, while their spatial (and kinematic) distributions provide insight into the structures in which the stars formed and their subsequent redistribution by mergers and secular processes. These observables are all coupled; studying the detailed relationships between them helps us identify typical scenarios and unusual cases. We can therefore hope to understand more about the developmental histories of galaxies and their internal components by studying the wavelength dependence of their spatial structure.

Galaxies can be split visually into two main morphological types: elliptical and spiral. The primary distinguishing feature is the presence of a disc component.  Lenticular galaxies are commonly grouped with ellipticals, as early-types, although their disc-component suggests they are more closely related to spirals \citep{VandenBergh1976}.  To complicate matters, even ellipticals sometimes appear to have a somewhat `disky' appearance \citep{Kormendy1996} or kinematics \citep{Cappellari2011}.

The radial luminosity profile of a galaxy, or its components, is often modelled by a \sersic profile.  This function provides a good description of how the surface brightness of many galaxies varies with radius \citep{Sersic1963,Graham2005}.

Elliptical galaxies are typically well-described by a single \sersic profile.  The index is often taken to be fixed at $n=4$ \citep{DeVaucouleurs1948}, although it has been shown to vary from $n \la 2$ at low masses to $n \ga 4$ for high mass ellipticals \citep{Hodge1971,Caon1993,Graham2013}.

Spiral galaxies, on the other hand, generally require a combination of two different \sersic profiles. The discs of spirals (and lenticulars) can usually be accurately described by an exponential ($n = 1$) profile (e.g., \citealt{Kormendy1977}, \citealt{Allen2006}). Bulges are often described by a de Vaucouleurs ($n = 4$) profile. As with elliptical galaxies, this is usually not accurate \citep{Andredakis1995,Graham2001,DeSouza2004}, so recent studies (including this one) tend to adopt a general \sersic profile, where $n$ is not fixed.  Some studies also account for the presence of a bar, commonly with a third \sersic profile \citep{Gadotti2011}.

Ideally we would be able to decompose galaxies into all their constituent structures.  However, with limited physical resolution and signal-to-noise, even bulge-disc decompositions can prove difficult to perform robustly. Although many galaxies comprise multiple components, fitting with a single \sersic profile can provide valuable insights into their structure, providing sizes, axial ratios, rough morphological classifications and total photometry (e.g., \citealt{Kelvin2012}).

Fitting \sersic models to galaxy images in different wavebands allows one to explore the dependence of galaxy structure on wavelength for large samples in a homogeneous manner.  Variation in a galaxy's surface brightness profile parameters with wavelength is qualitatively equivalent to a radial trend in the colour of the galaxy.  Such `colour gradients' have been well-studied, usually by measuring the surface brightness in a series of elliptical annuli.  This approach is usually limited by the seeing, and hence confined to relatively extended objects.  The result is a colour profile for the galaxy, usually summarised as the overall change in colour over some (logarithmic) radial range (e.g., \citealt{Goudfrooij1994, DeJong1996, Saglia2000, denBrok2011}). 

The wavelength-dependence of \sersic model parameters provides an alternative description.
This is less detailed than a full colour profile, though more robustly determined in typical survey imaging.
In addition, as the PSF can be easily accounted for when fitting the model, the results are not biased by seeing.

Considering the wavelength-dependence of \sersic parameters is more efficient than dealing with empirical colour profiles, while providing significantly more general information than quoting a colour gradient over a single (or several) radial ranges.  Furthermore, the ubiquity of the \sersic profile in describing galaxy profiles, and the correlations between its parameters and other galaxy properties, suggests that the profile and its parameters have a physical significance.
Expressing colour gradients in terms of the wavelength-dependence of effective (half-light) radius and \sersic index naturally separates the proportions associated with changes in size and profile shape, which may have different physical drivers.  However, the price of all this is the assumption of a particular functional form for the surface brightness profile.  When this assumption is inappropriate (e.g., in merging galaxies, multi-component systems or potentially very high redshift), it may result in biases or misinterpretations.

Some studies have chosen a compromise between performing a parametric fit and directly measuring a colour profile. \citet{LaBarbera2009} fit independent \sersic models in various wavebands and then use elliptical annuli to measure colour profiles and gradients on the model images.  This removes the effect of the PSF and reduces noise in the measurements.

Several groups have examined the trends of \sersic model parameters with wavelength.  For a sample of bright elliptical galaxies, \citet{LaBarbera2010b} find that the mean effective radius decreases significantly with increasing wavelength from $g$ to $H$, but that there is little variation in \sersic index.
\citet{Kelvin2012} find similar behaviour in data from the Galaxy And Mass Assembly survey (GAMA; \citealt{Liske2015}).
They also consider late-types, finding a substantial increase in \sersic index, and decrease in effective radius, across the same wavelength range.  Metallicity gradients and dust attenuation are proposed as reasons for these trends.

\citet{Lange2015} have recently extended the work of \citet{Kelvin2012}, replacing the GAMA infrared UKIDSS LAS imaging \citep{Lawrence2007} with deeper VISTA-VIKING data \citep{Edge2013}. They also separate the galaxy population into early- and late-types using a variety of dividers, including visual classification.  \citeauthor{Lange2015} find that their results are not sensitive to chosen divider, but show that the bimodality of galaxy structure becomes less distinct at lower stellar masses. They confirm that galaxies appear more compact at redder wavelengths, although this appears less dramatic than in the previously mentioned studies.
\citeauthor{Lange2015} argue that these structural variations with wavelength may arise from the two-component nature of many galaxies, in which the bulge is observed in the redder wavebands whilst the disc is observed in the bluer wavebands (also see \citealt{Driver2007a}).
However, they do not examine the dependence of \sersic index on waveband.

The above studies fit the image in each filter-band independently.  This model freedom can be seen as an advantage.  However, such an approach does not utilise the expected strong correlations between structural parameters at neighbouring wavelengths. In data with low signal-to-noise or poor resolution, using these physical expectations can improve the reliability and precision of structural measurements.  Furthermore, even in high-quality data, utilising available colour information can improve the performance of decompositions.  To address these issues, the \megamorph project \citep{Bamford2012} has developed a technique which fits a single, wavelength-dependent model simultaneously to a set of imaging in different filter-bands. These developments have been implemented in an extended version of \galfit \citep{Peng2002,Peng:2009fu} named \galfitm (\citealt{Haussler2013}, hereafter H13; \citealt{Vika2013,Vika2014}).  \galapagos \citep{Barden2012}, a software package dedicated to running \galfit on large surveys, has been similarly extended to become \galapagosII, which uses \galfitm (H13).

Using \megamorph multi-band techniques to perform single-\sersic fits, \citeauthor{Vulcani2014} (2014, hereafter V14) studied the wavelength dependence of galaxy effective radius and \sersic index. They confirmed the trends described above, and demonstrated that galaxies with contrasting total colours and \sersic indices display strikingly different behaviour in terms of the wavelength dependence of their effective radii and \sersic indices.

V14 speculate that the behaviour seen for early-types may result from the superposition of red compact structures and more-extended, bluer structures, all of which possess the $n \sim 4$ profile of systems supported by random motions.  This is similar to the picture presented by \citet{Huang2013}. The behaviour of late-types is attributed to their two-component nature, as mentioned above.

In this paper we aim to gain a greater confidence in, and deeper understanding of, the trends of galaxy structural parameters with wavelength as measured by the various studies described above.  This work will advise our application of multi-band structural fitting to the full set of high-quality imaging currently being assembled by GAMA, and our subsequent analysis.
We build on the results of V14, first ensuring they are robust to redshift effects.

For the first time, we examine the wavelength-dependence of galaxy structure as a function of luminosity.  We also show that the measured trends persist for selections based on visually classifed morphology, rather than morphological proxies.  Next, we investigate the role of dust in driving the wavelength dependence of late-type galaxy structure.  We conclude by discussing the various potential factors responsible for the variation of galaxy structural parameters with wavelength.  By taking the novel approach of considering the joint wavelength dependence of \sersic index and effective radius, we can begin to decouple the relationships between concentration, size and colour.

The analysis has been carried out using a cosmology with ($\Omega_{m},\Omega_{\Lambda}, h) = (0.3, 0.7, 0.7)$ and AB magnitudes.

\section{Data}

\subsection{Parent sample and structural parameters}

The galaxy measurements used in this paper have previously been presented in H13, and a subsample of these were studied in V14. A detailed description of the selection criteria, robustness of fits and properties of the sample can be found in those papers; here, we only give a brief overview.

Our data is taken from the extended Galaxy And Mass Assembly survey  (GAMA II) (Liske et al, submitted), which is the largest homogeneous, multi-wavelength dataset currently available for the low-redshift Universe.  The imaging data GAMA has assembled, from both SDSS (\citealt{York2000}) and UKIDSS (\citealt{Lawrence2007}), provides a consistent set of pixel-registered multi-wavelength data, covering the ($ugriz$) optical bands and the ($YJHK$) near-infrared (NIR) bands, respectively \citep{Hill2011}. It has been demonstrated that these data have sufficient depth and resolution to allow \sersic profiles to be fit to large samples (\citealt{Kelvin2012}; H13).

As in V14, the data we use in this paper are limited to the G09 region of GAMA. We therefore only utilise approximately one-fifth of the area covered by GAMA II.  For efficiency, we chose to focus on a single region for our initial exploration of multi-band fitting techniques with GAMA.  The application of \megamorph methods to the full GAMA dataset, with UKIDSS-LAS replaced by VISTA-VIKING data, are ongoing.

We use \textsc{galapagos-2} with \textsc{galfitm} to fit a single wavelength-dependent 2D model to all images of a galaxy simultaneously. \galfitm allows each parameter in the standard \sersic light profile to vary as a function of wavelength; the user may choose the degree of smoothness they require. In this work, galaxy magnitudes are allowed to vary freely, while \sersic index and effective radius are modelled as quadratic functions of wavelength.  All other parameters are not permitted to vary with wavelength.  \galfitm accounts for seeing by convolving the model with a PSF before comparing to the data.  We utilise the PSFs produced by \citet{Kelvin2012}.

Restframe values of \sersic index and effective radius for each wavelength have been obtained from the polynomials returned by \galfitm.  For the magnitudes, k-corrections have been performed using InterRest (\citealt{Taylor2009}).

\subsection{Sample selection}

We take the \galapagosII output catalogue, however not every fit was necessarily accurate or meaningful, so a number of cleaning criteria were applied to eliminate galaxies for which the fit has violated our criteria in one or more bands. These criteria are identical to those used in H13 and V14, with the exception of \sersic index which is limited to $0.201 < n < 7.75$.  We have reduced the upper limit on \sersic index to avoid including suspected poor fits, with final parameters close to the fitting constraint (objects with very high $n$ are typically poor fits, often to point sources, and are therefore removed).

\begin{figure}

	\includegraphics[width=0.45\textwidth]{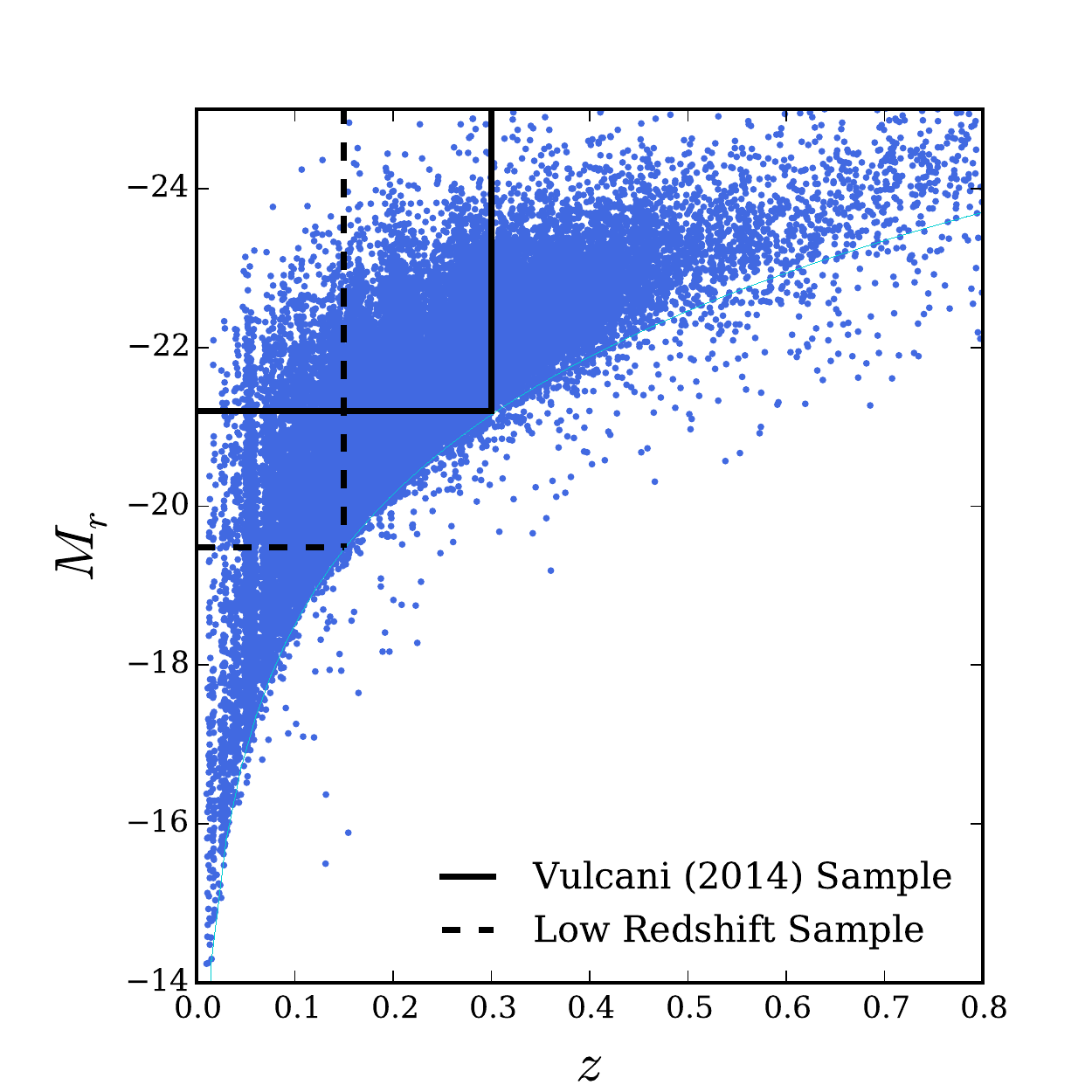}
	\caption{Absolute $r$-band magnitude versus redshift for our parent sample, with our volume-limited selection boxes overlaid.  The curve line indicates the primary apparent magnitude limit of the GAMA II redshift survey, $r < 19.8$. This corresponds to an absolute magnitude of $M_{r} = -21.2$ at $z = 0.3$ and $M_{r} = -19.48$ at $z = 0.15$.
	\label{fig:Vol_lim}}
\end{figure}

In this paper we use two different volume-limited samples to avoid selection effects, which are illustrated in Fig.~\ref{fig:Vol_lim}. For studying variation in galaxy properties with redshift, a first volume-limited sample of 13,871 galaxies is taken with $z < 0.3$, $M_{r} < -21.2$, in line with V14. Within this redshift limit the restframe $H$-band can be interpolated from measurements in the observed $u$--$K$ wavelength range.

For studying variation in galaxy properties with absolute magnitude, $M_{r}$, a second volume-limited sample ($z < 0.15$, $M_{r} < -19.48$) of 5331 galaxies is used, allowing galaxies to be considered over a wider range of absolute magnitude.

Note that the cleaning described above removes only 0.42 / 0.39 per cent of our V14 / low-redshift samples.

One of the aims of this project is to study how galaxy properties vary as a function of wavelength. It is expected that different types of galaxies will behave in different ways, so we subdivide our sample (as in V14) by both colour and \sersic index. The number of galaxies in each subsample are given in Tables \ref{table:Vulcani_sample} \& \ref{table:Low_z_sample} for both volume-limited samples, after basic cleaning has been applied.

\begin{table}
\centering

	\begin{tabular}{| l | l | l | l | l | l |}
		\hline
		 & & \multicolumn{2}{ c|}{$n_{r} < 2.5$} & \multicolumn{2}{ c|}{$n_{r} > 2.5$}  \\ \hline
        & \multicolumn{1}{|c|}{Colour} & no. & \% & no. & \% \\ \hline
		\multicolumn{1}{| l }{Blue:}  &  \multicolumn{1}{ r|}{$(u-r) < 1.6$}  & \multicolumn{1}{ l }{1380} & \multicolumn{1}{ l|}{9.9}  & \multicolumn{1}{ l }{322} & \multicolumn{1}{ l|}{2.3} \\ \hline
		\multicolumn{1}{| l }{Green:} & $1.6 < (u-r) < 2.1$ & \multicolumn{1}{ l }{3311} &  \multicolumn{1}{ l|}{23.9} & \multicolumn{1}{ l }{1047} &  \multicolumn{1}{ l|}{7.5} \\ \hline
		\multicolumn{1}{| l }{Red:}   &   $2.1 < (u-r)$  & \multicolumn{1}{ l }{2523} &  \multicolumn{1}{ l|}{18.2} & \multicolumn{1}{ l }{5288} &  \multicolumn{1}{ l|}{38.1} \\ \hline
	\end{tabular}
	\caption{Number count and fraction of the $z < 0.3$, $M_{r} < -21.2$ volume-limited sample (total = 13871 galaxies), for different combinations of colour and \sersic index.
	\label{table:Vulcani_sample}}
\end{table}

\begin{table}
\centering

	\begin{tabular}{| l | l | l | l | l | l |}
		\hline
		& & \multicolumn{2}{|c|}{$n_{r} < 2.5$} & \multicolumn{2}{|c|}{$n_{r} > 2.5$}  \\ \hline
         & \multicolumn{1}{|c|}{Colour} & no. & \% & no. & \% \\ \hline
		\multicolumn{1}{| l }{Blue:}  &  \multicolumn{1}{ r|}{$(u-r) < 1.6$}  & \multicolumn{1}{ l }{1323} & \multicolumn{1}{ r|}{24.8}  & \multicolumn{1}{ l }{134} & \multicolumn{1}{ r|}{2.5} \\ \hline
		\multicolumn{1}{| l }{Green:} & $1.6 < (u-r) < 2.1$ & \multicolumn{1}{ l }{1417} &  \multicolumn{1}{ r|}{26.6} & \multicolumn{1}{ l }{245} &  \multicolumn{1}{ r|}{4.6} \\ \hline
		\multicolumn{1}{| l }{Red:}   &   $2.1 < (u-r)$  & \multicolumn{1}{ l }{994} &  \multicolumn{1}{ r|}{18.6} & \multicolumn{1}{ l }{1218} &  \multicolumn{1}{ r|}{22.8} \\ \hline
	\end{tabular}
	\caption{Number count and fraction of the $z < 0.15$, $M_{r} < -19.48$ volume-limited sample (total = 5331 galaxies), for different combinations of colour and \sersic index.
	\label{table:Low_z_sample}}
\end{table}

\begin{figure}
		\centering
		\includegraphics[width=0.5\textwidth]{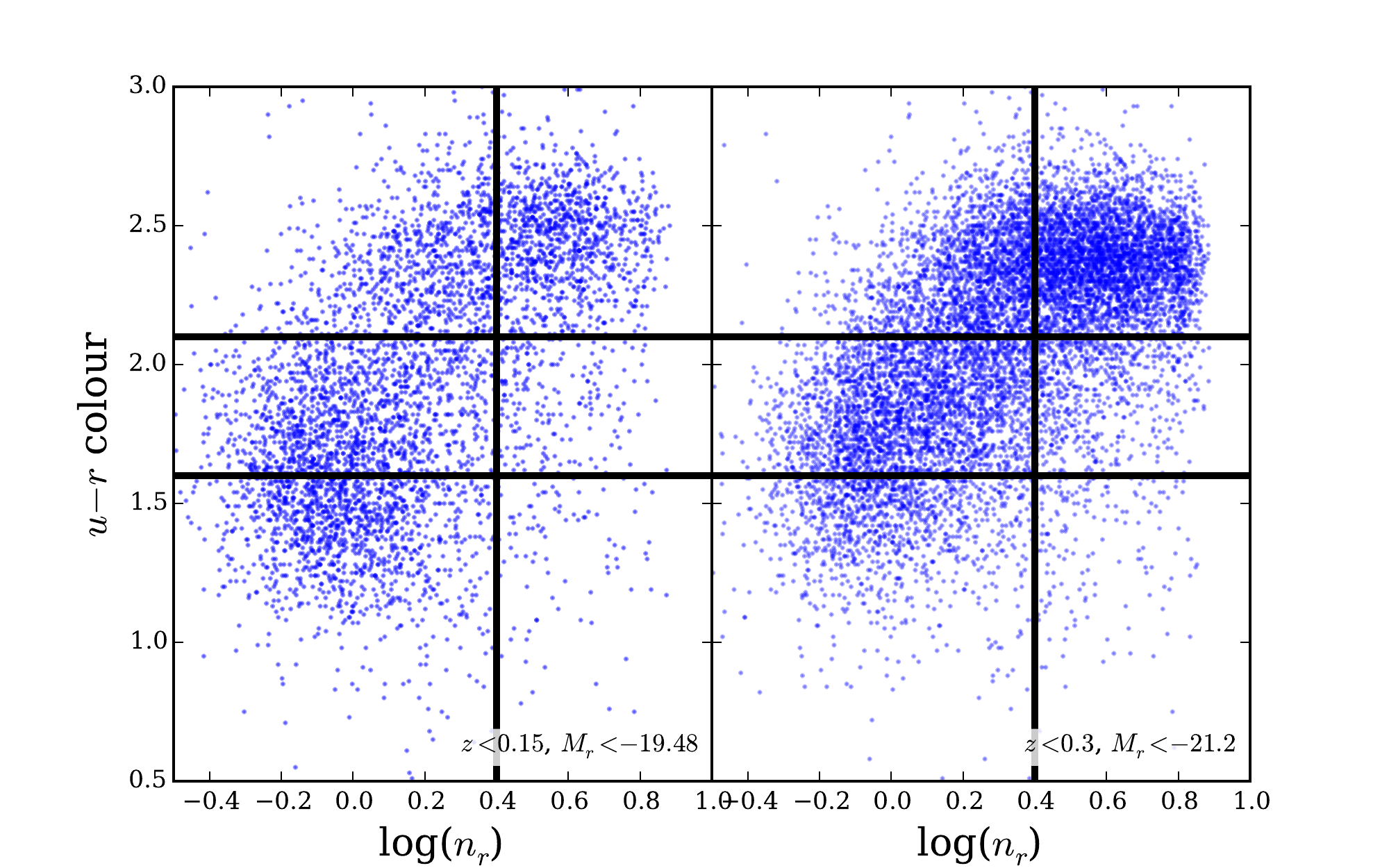}
				
		\caption{(u-r) rest frame colour vs. $n_{r}$ for the galaxies in our two samples. Lines illustrate the cuts we apply to divide the galaxies by colour and \sersic index.
		\label{fig:n_colour}}
		
	\end{figure}

Initially the samples are divided into `red' and `blue' at $(u-r) = 2.1$. We also want to separate the bluest galaxies, which may contain starbursts, so we further divide this `blue' sample at $(u-r) = 1.6$ into `green' and `blue'.  Note that our `green' sample corresponds to the main population of star-forming galaxies, not the green valley.  \B{We have confirmed that altering these colour cuts does not affect our results.}  The galaxy sample is also split by \sersic index, in an attempt to separate discy galaxies from ellipticals, at $n_r = 2.5$. These divisions can be seen in Fig. \ref{fig:n_colour}.

\section{Results}

V14 examined the variation of $n$ and $\re$ with wavelength in order to reveal trends of internal structure, and therefore formation history, for galaxies in different subsamples. For consistency we continue to use the notation $\mathcal{N}^{x}_{g} = n(x)/n(g)$ and $\mathcal{R}^{x}_{g} = \re(x)/\re(g)$, which express the variation for a given waveband, $x$, versus that in the $g$-band. For simplicity, we will only consider the ratio of $H$-band to $g$-band in this paper, as V14 shows that there is a consistent trend with wavelength and this pair has the longest robust wavelength baseline.  We omit the band labels from \N and \R when discussing their general behaviour.

If we convert our \N and \R trends to colour gradients we find good agreement with previous work, e.g. $red$, high-$n$ galaxies show identical colour gradients to the passive, `early-type' sample studied in \cite{LaBarbera2010}.  We also see the same behaviour of colour gradient with luminosity. However, here we focus on \N and \R, which allows us to decouple variations in concentration and size with wavelength, which are combined in a measurement of colour gradient alone.

V14 concluded that, regardless of overall $u-r$ colour, low-$n$ galaxies display a very different behaviour to high-$n$ galaxies, implying that both \N and \R depend strongly on \sersic index. They suggested that low-$n$ discy (two-component) galaxies show a large change in \sersic index with wavelength because in bluer wavebands the \sersic index of the disc is being measured, whereas in redder wavebands the \sersic index of the bulge dominates. High-$n$ galaxies are, on the other hand, likely to be bulge-dominated (closer to one-component), so there is little change in \sersic index with wavelength. There is, however, a large change in $\re$ with wavelength for high-$n$ galaxies, which indicates that elliptical galaxies contain a number of different pressure-supported stellar populations with different extent, possibly resulting from multiple merging events throughout a galaxy's lifetime.

\subsection{Redshift effects on galaxy structure}
\label{subsec:Galaxy_Z}

The sample used in V14 extends to $M_{r} < -21.2$, $z < 0.3$, so much of the sample is faint and of small angular size.  This selection was partly chosen to demonstrate the power of multi-band fitting in this challenging regime. V14 showed that the measured \N and \R correspond to differences in the visual appearance of the galaxies. However, before exploring the luminosity dependence of these trends, we will further test the resilience of our \N[H][g] and \R[H][g] measurements versus redshift.

\begin{figure}
		\centering
		\includegraphics[width=0.45\textwidth]{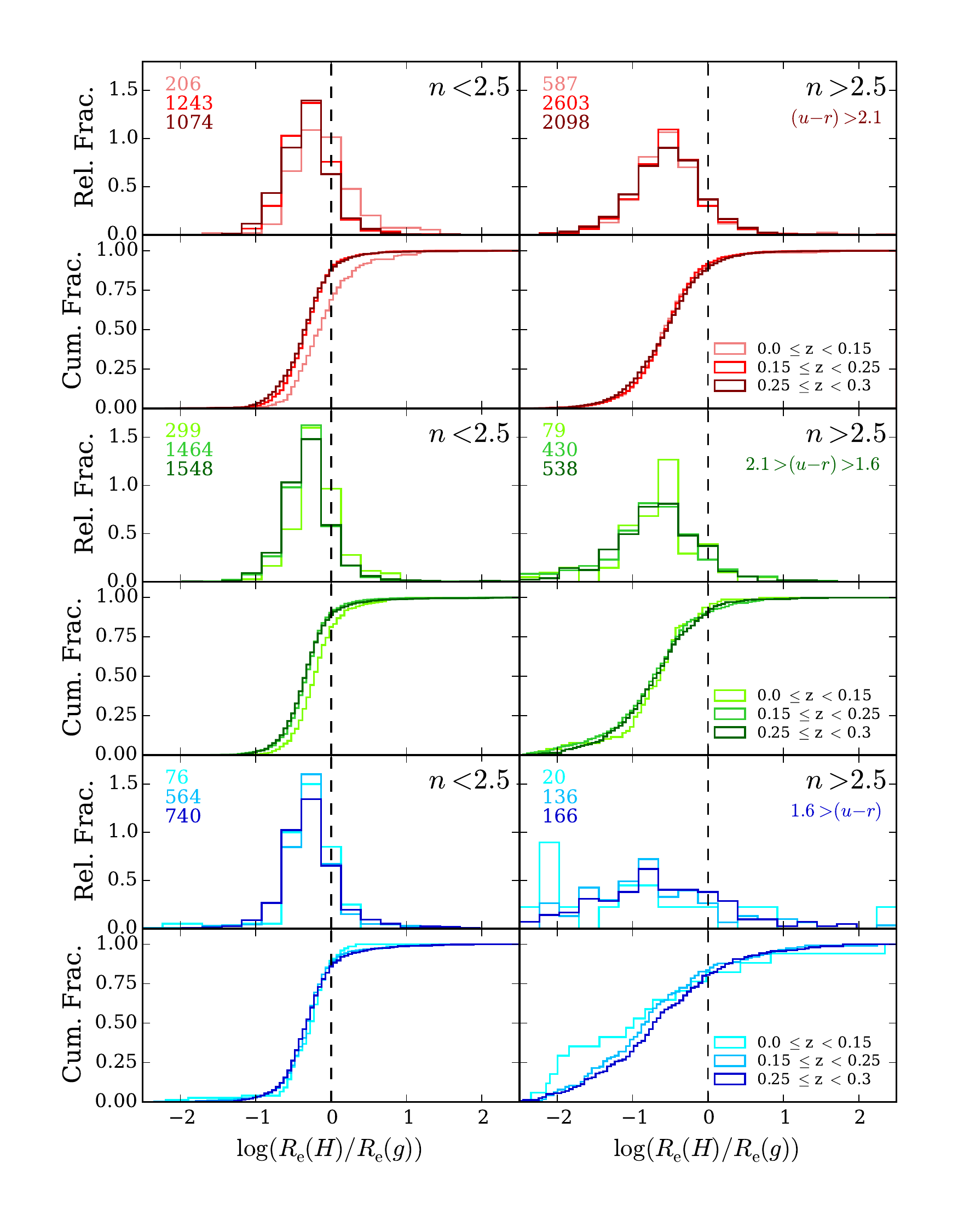}
				
		\caption{Effective radii of galaxies binned by redshift, for colour and \sersic index subsamples. The left column shows low-$n$ galaxies, the right column shows high-$n$ galaxies.  The top, middle and bottom groups of four panels contain \emph{red}, \emph{green} and \emph{blue} samples, respectively. Within each group, the top panels show normalised histograms, whilst the bottom panels present corresponding cumulative histograms to better visualise subtle shifts in the distributions. In each panel lines in light colours show galaxies in the low redshift bin ($0 \leq z < 0.15$), mid-shades indicate the intermediate redshift bin ($0.15 \leq z < 0.25$), whilst  the darkest colours show galaxies in the high redshift bin ($0.25 \leq z < 0.3$).
		\label{fig:Galaxies_Re_Z}}
		
		\end{figure}

We split our bright volume-limited sample ($z < 0.3$, $M_{r} < -21.2$) into three redshift bins: $0.0 \leq z < 0.15$, $0.15 \leq z < 0.25$ and $0.25 \leq z < 0.3$. These bins were chosen to span the redshift range of our volume-limited sample, while ensuring a similar number of galaxies in each bin to permit meaningful comparisons (see Table \ref{table:Vulcani_sample}).
In Fig. \ref{fig:Galaxies_Re_Z} we show the wavelength dependence of effective radius, for galaxies split by $u-r$ colour, $n_{r}$, and redshift.  Galaxies of all colours and \sersic indices are typically found to have \R[H][g] $< 1$, indicating they are smaller at redder wavelengths, while $\mathcal{R}^{H}_{g} = 1$ (i.e. $\log{\mathcal{R}^{H}_{g}}=0$) corresponds to no variation in size with wavelength.  Here, \R[H][g] does not appear to change substantially with redshift in any subsample. Kolmogorov--Smirnov (KS) tests do indicate some significant differences between the \R[H][g] distributions in the different redshift samples.  However, these differences are generally small, particularly in comparison with the width of each distribution and the offsets between the low- and high-$n$ subsamples.  To determine whether an offset between subsamples can be considered `large' or `small' we sum the standard deviations of the widest and narrowest distributions in quadrature. We then find the difference in the median value of $log(\mathcal{R}^{H}_{g})$ in the highest and lowest redshift bins, as a fraction of the summed standard deviation. Here, there average offset is 15.5\% of the distribution widths, which can be considered small.

What trends are present, indicate that at lower redshift (higher S/N and better resolution) we measure \R[H][g] closer to unity for low-$n$ galaxies. The departure of \R[H][g] from unity may therefore be slightly overestimated in V14. High-$n$ systems appear unaffected, with perhaps the exception of blue, high-$n$ galaxies. There are, however, very few galaxies in this subsample as blue spheroids are uncommon.  We therefore consider our measurements of the wavelength dependence of effective radius to be robust out to $z = 0.3$.

\begin{figure}
		\centering
		\includegraphics[width=0.45\textwidth]{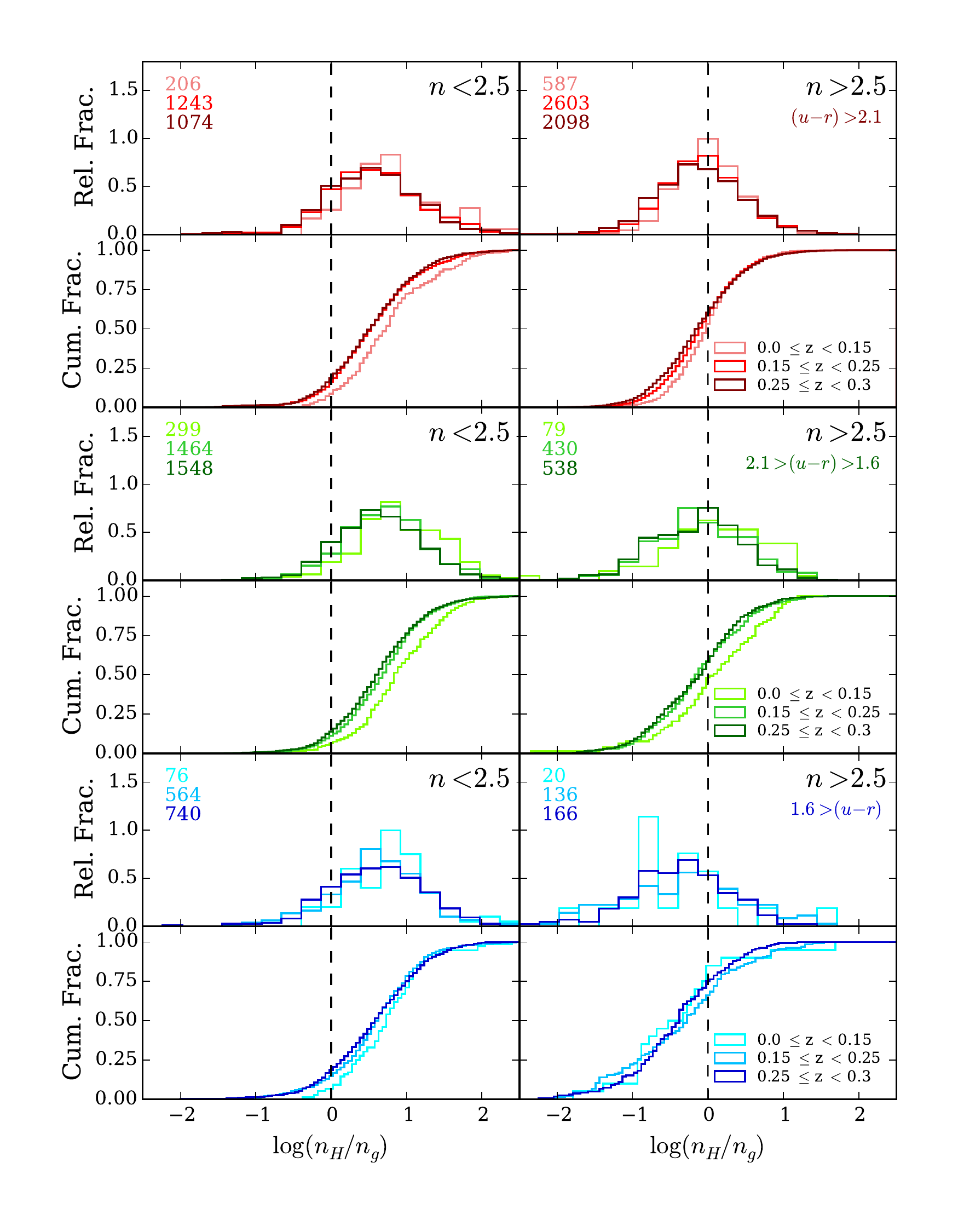}		
		
		\caption{\sersic indices of galaxies binned by redshift, for colour and \sersic index subsamples.  Panels and colours are chosen similar to Fig \ref{fig:Galaxies_Re_Z}.
		\label{fig:Galaxies_N_Z}}
		
		\end{figure}

We study \N[H][g] in the same manner.  In Fig. \ref{fig:Galaxies_N_Z} we show the wavelength dependence of \sersic index for galaxies split by $u-r$ colour, $n_{r}$, and redshift bin. Again, some small trends are visible. Galaxies at lower-$z$ tend to display \N[H][g] further from unity. This is most significant for low-$n$ galaxies, and suggests that V14 may slightly underestimate the wavelength dependence of $n$ for these systems.  Apart from this small effect (approx. 18.5\% offset in median values compared to the width of the distributions, as described previously), we consider our measurements of the wavelength dependence of \sersic index to be robust out to $z = 0.3$.

\begin{figure}
	\centering
	\includegraphics[width=0.45\textwidth]{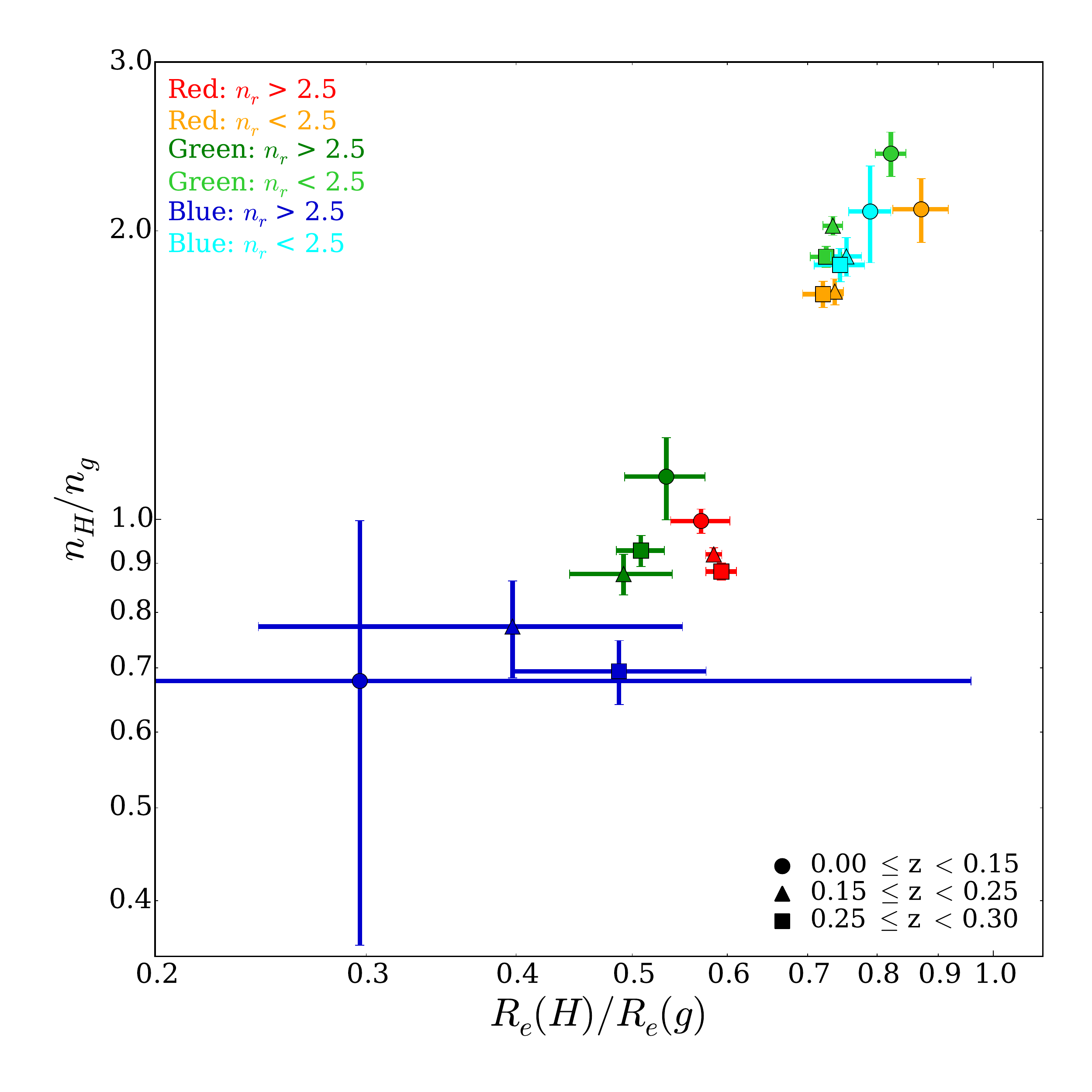}
	\caption{Median	$\mathcal{N}^{H}_{g}$ versus median $\mathcal{R}^{H}_{g}$ for galaxies in each of our different subsamples divided by colour, \sersic index and redshift bin.  Colours are the same as in figure 17 of V14.  Error bars show uncertainty on the median, estimated as $1.253\sigma/\sqrt{N}$, where $\sigma$ is the standard deviation about the median and $N$ is the number of galaxies in the sample.  There are small trends with redshift due to observational biases.  However, the distinction between the subsamples, particularly the contrast between low- and high-$n$, remains very clear.  Note that the \emph{blue} $0 \leq z < 0.15$, $n > 2.5$ point lies at $\mathcal{R}^{H}_{g} =  0.15$, $\mathcal{N}^{H}_{g} = 0.56$, but is consistent with the higher $z$ points given its large uncertainties.
	\label{fig:Vulcani_01}}
\end{figure}

As in Fig. 17 of V14, we summarise the relationship between \N[H][g] and \R[H][g] for our three redshift bins in Fig.~\ref{fig:Vulcani_01}. The small trends with redshift, described above, are again evident in this figure.  However, for each subsample, the redshift bins lie close to one another.  In addition to observational effects, redshift trends may result from real changes in the galaxy population.  However, given our volume-limited, colour-selected samples, and the narrow range of redshift, we do not expect evolution of the galaxy population to play a significant role in these trends.  The distinctions between subsamples, particularly high- and low-$n$ systems, are maintained independent of redshift. The conclusions of V14 are thus robust.  With that established, we now move on to examining how the wavelength dependence of galaxy structure varies with luminosity.

\subsection{Luminosity-dependence of galaxy structure}
\label{subsec:Galaxy_Mr}

V14 specifically chose a volume-limited sample of bright galaxies with $M_{r} < -21.2$. Here we investigate whether the conclusions drawn from their study apply to a wider range of galaxy luminosities.

Our sample is divided into three magnitude bins:  $-22.48 \leq M_{r} < -21.48$, $-21.48 \leq M_{r} < -20.48$ and $-20.48 \leq M_{r} < -19.48$. These intervals ensure similar numbers of galaxies in each bin, allowing for meaningful comparisons.

\begin{figure}
		\centering
        \includegraphics[width=0.45\textwidth]{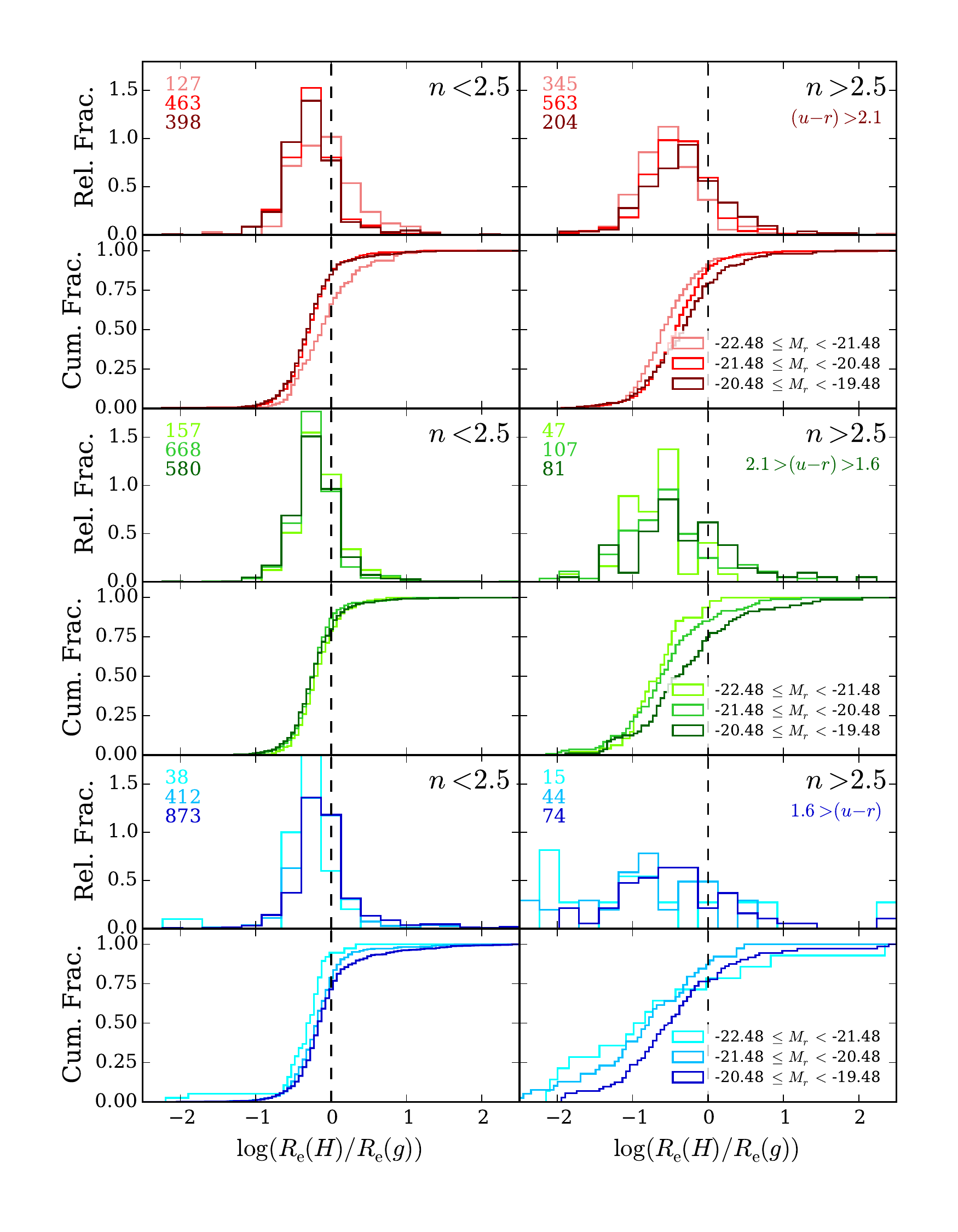}
		
		\caption{Effective radii of galaxies binned by absolute magnitude, for colour and \sersic index subsamples.  Panels and colours are chosen similar to Fig \ref{fig:Galaxies_Re_Z}.
		\label{fig:Galaxies_Re_Mr}}
		
		\end{figure}

In Fig.~\ref{fig:Galaxies_Re_Mr} we present the wavelength dependence of effective radius for subsamples divided by $u-r$ colour, $n_{r}$ and luminosity.

For high-$n$ galaxies, there appears to be a consistent trend in the distribution of \R[H][g] with changing absolute magnitude. Fainter high-$n$ galaxies have \R[H][g] closer to unity than brighter high-$n$ galaxies.  Their sizes therefore vary less as a function of wavelength.  The most luminous galaxies, such as those studied by V14, display the strongest variation of their sizes with wavelength.

For the low-$n$ samples, although K-S tests indicate some differences between the distributions in the luminosity bins, there is little obvious consistent variation.  The \emph{blue} sample displays similar trends to the high-$n$ galaxies, while typical \emph{green} galaxies show no luminosity dependence. \emph{Red} galaxies hint at an opposing behaviour, such that the most luminous bin displays less variation in size with wavelength.

On average across all samples, the offset of the distributions with respect to the standard deviation of the distributions is 27\%, calculated as above, for the redshift distributions.

\begin{figure}
	\centering
	\includegraphics[width=0.45\textwidth]{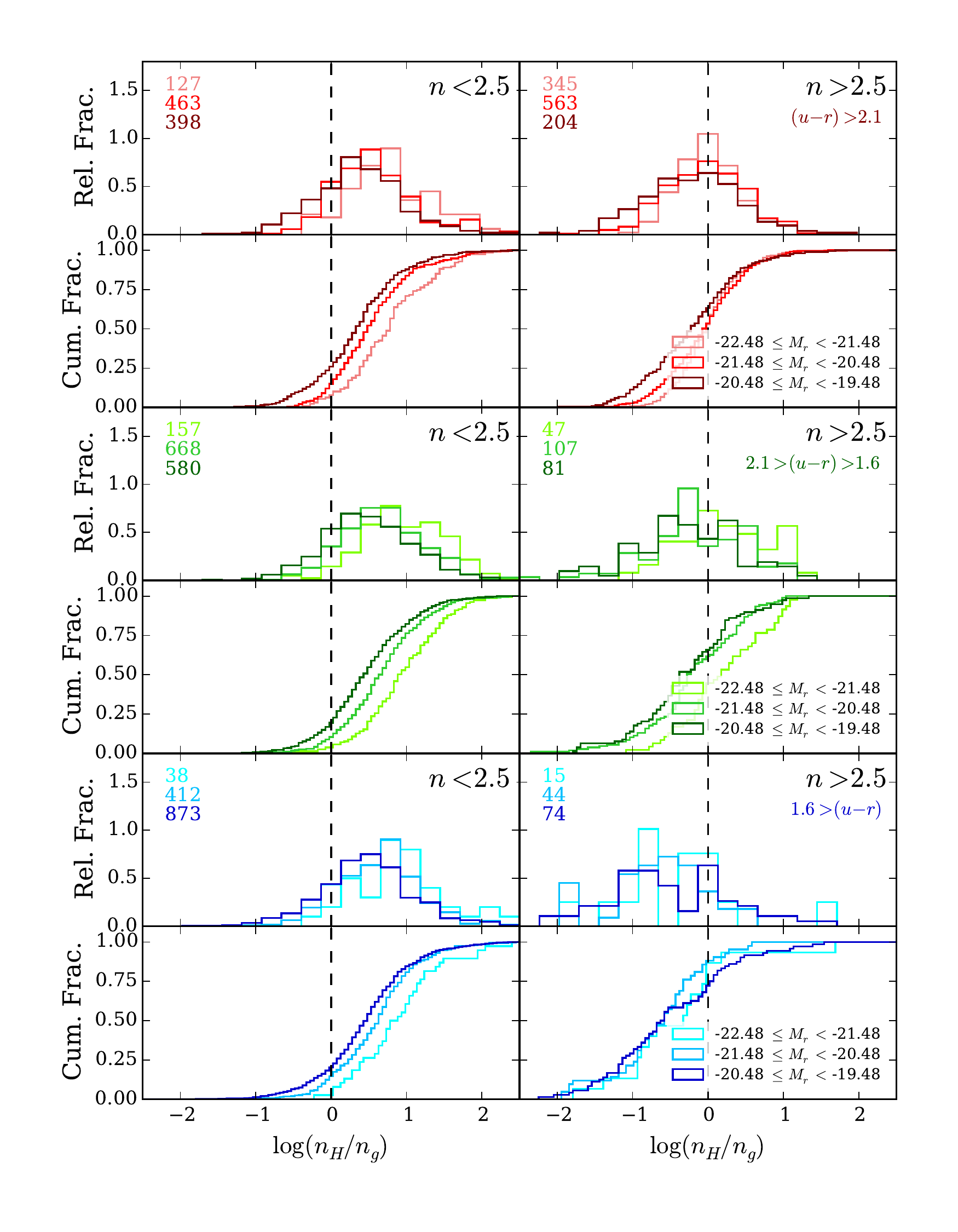}
	
	\caption{\sersic index of galaxies binned by absolute magnitude, for colour and \sersic index subsamples.  Panels and colours are chosen similar to Fig \ref{fig:Galaxies_Re_Z}.
	\label{fig:Galaxies_N_Mr}}
	
	\end{figure}

Again we study \N[H][g] in the same manner; Fig.~\ref{fig:Galaxies_N_Mr} shows the wavelength dependence of \sersic index for the same subsamples as Fig.~\ref{fig:Galaxies_Re_Mr}. There appears to be a consistent trend in the distribution of \N[H][g] with changing absolute magnitude for both low- and high-$n$ galaxies, with a 47\% mean offset of the distributions, compared to the width of those distributions. In the high-$n$ sample, distributions for all luminosity bins peak around one, supporting the conclusion of V14 that bulge-dominated (and likely one-component) systems show little change in \sersic index with wavelength. Comparing cumulative distributions for the high-$n$ samples, we see that bluer and less luminous samples tend to have slightly lower values of \N[H][g].  Typical bright high-$n$ galaxies have \N[H][g] very close to unity.

The low-$n$ samples consistently display values of \N[H][g] above one, indicating an increase in \sersic index with wavelength.  This offset from unity strongly increases with luminosity, such that the brightest low-$n$ galaxies display the greatest dependence of \sersic index on wavelength.

\begin{figure}
	\centering
	\includegraphics[width=0.45\textwidth]{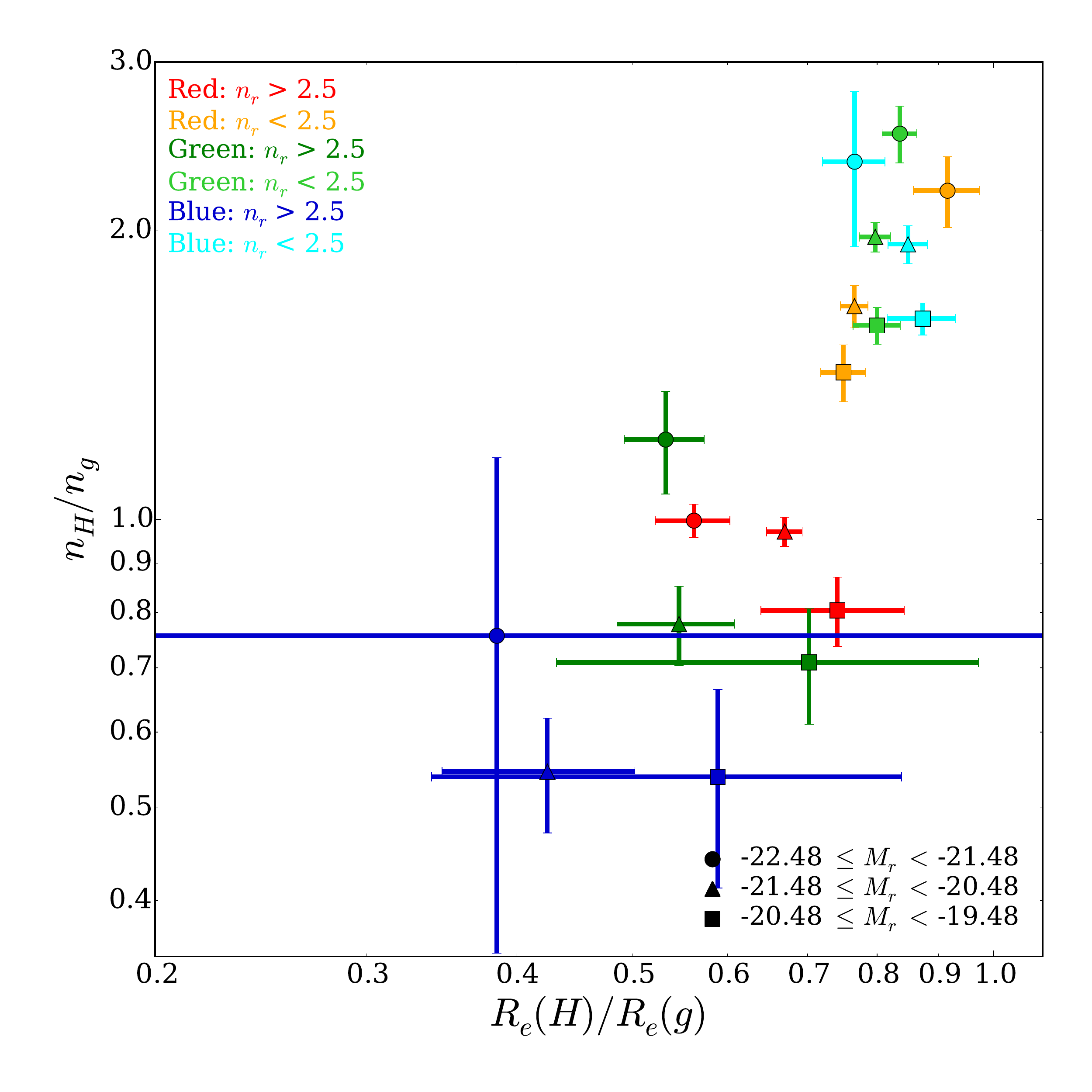}
	\caption{Median	\N[H][g] versus median \R[H][g] for galaxies in each of our different subsamples divided by colour, \sersic index and luminosity. Colours are the same as in figure 17 of V14.
	\label{fig:Vulcani_02}}
\end{figure}

In a similar manner to before, Fig.~\ref{fig:Vulcani_02} shows the relationship between \N[H][g] and \R[H][g] for our three magnitude bins.  High \sersic index galaxies show a clear variation in effective radius with luminosity (brighter galaxies showing a greater decrease in $\re$ with wavelength). Low \sersic index galaxies show a change in $n$ with wavelength (brighter galaxies showing a greater increase in $n$ with wavelength).  The overall effect is that the differences between low- and high-$n$ galaxies become more pronounced with increasing luminosity.

We have performed all of the analysis in Sec.~\ref{subsec:Galaxy_Z} and \ref{subsec:Galaxy_Mr} using an optical baseline ($u$--$z$), although using the same fits to the full $u$--$K$ dataset. As one would expect, over this narrower wavelength range the differences between the samples narrow, though not in terms of significance, as the scatter shrinks roughly in proportion.  All of the behaviour seen in the $g$ versus $H$ plots is qualitatively the same, and our inferences would be unchanged.

In terms of overall colour, concentration and size, early- and late-type galaxies become more similar with increasing luminosity. However, when one considers the joint distribution of these three properties, as in this paper, one instead finds a divergence which widens with luminosity. The process by which late-type (low-$n$) galaxies grow must promote the variation of profile shape with wavelength.  The growth of early-type (high-$n$) galaxies must also maintain a radial segregation of their stellar populations, but crucially in terms of size, not profile shape.  Furthermore, any process which transforms late-types to early-types must do so in a manner that erases the \B{wavelength-variation} of profile shape, but enhances that of size.  Realistic mechanisms for galaxy growth and transformation must reproduce the trends we observe.

\subsection{Morphological classifications}

\begin{figure}
	\centering
	\includegraphics[width=0.45\textwidth]{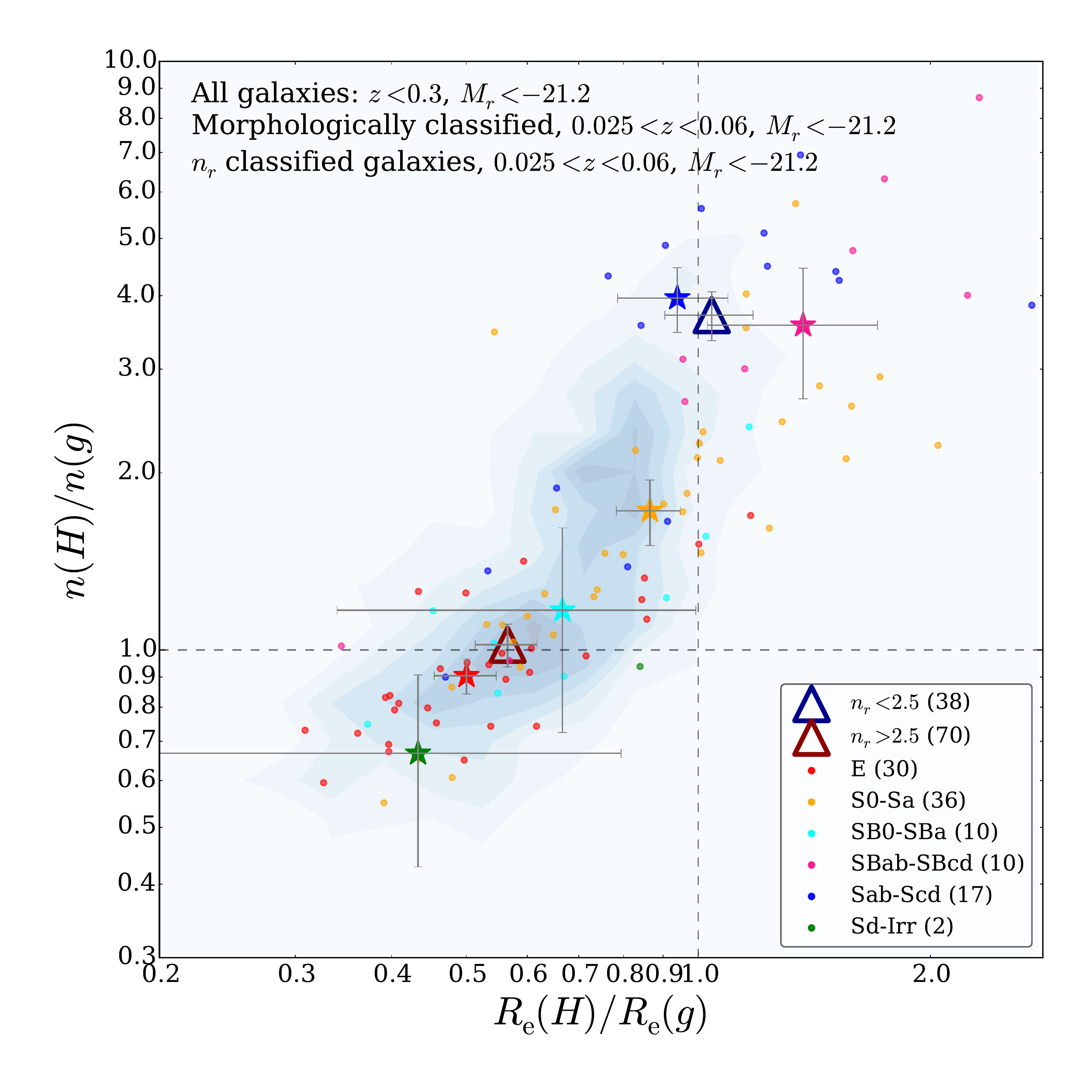}
	\caption{\N[H][g] versus \R[H][g] for galaxies in our high-luminosity volume-limited sample with Hubble type classifications in the GAMA visual morphology catalogue. Median points for each Hubble type bin are plotted as stars, with associated error bars (estimated as $1.253\sigma/\sqrt{N}$). Morphological classifications are only available for a low-redshift subset of galaxies.  For comparison, the median values for galaxy samples with the same redshift limits, but separated by $n_{r}$, are plotted as triangles. The distribution of the full volume-limited sample (to the same luminosity limit, but extending out further in redshift) is plotted as contours.
	\label{fig:Morph_Vulcani}}
\end{figure}

\begin{figure}
	\centering
	\includegraphics[width=0.45\textwidth]{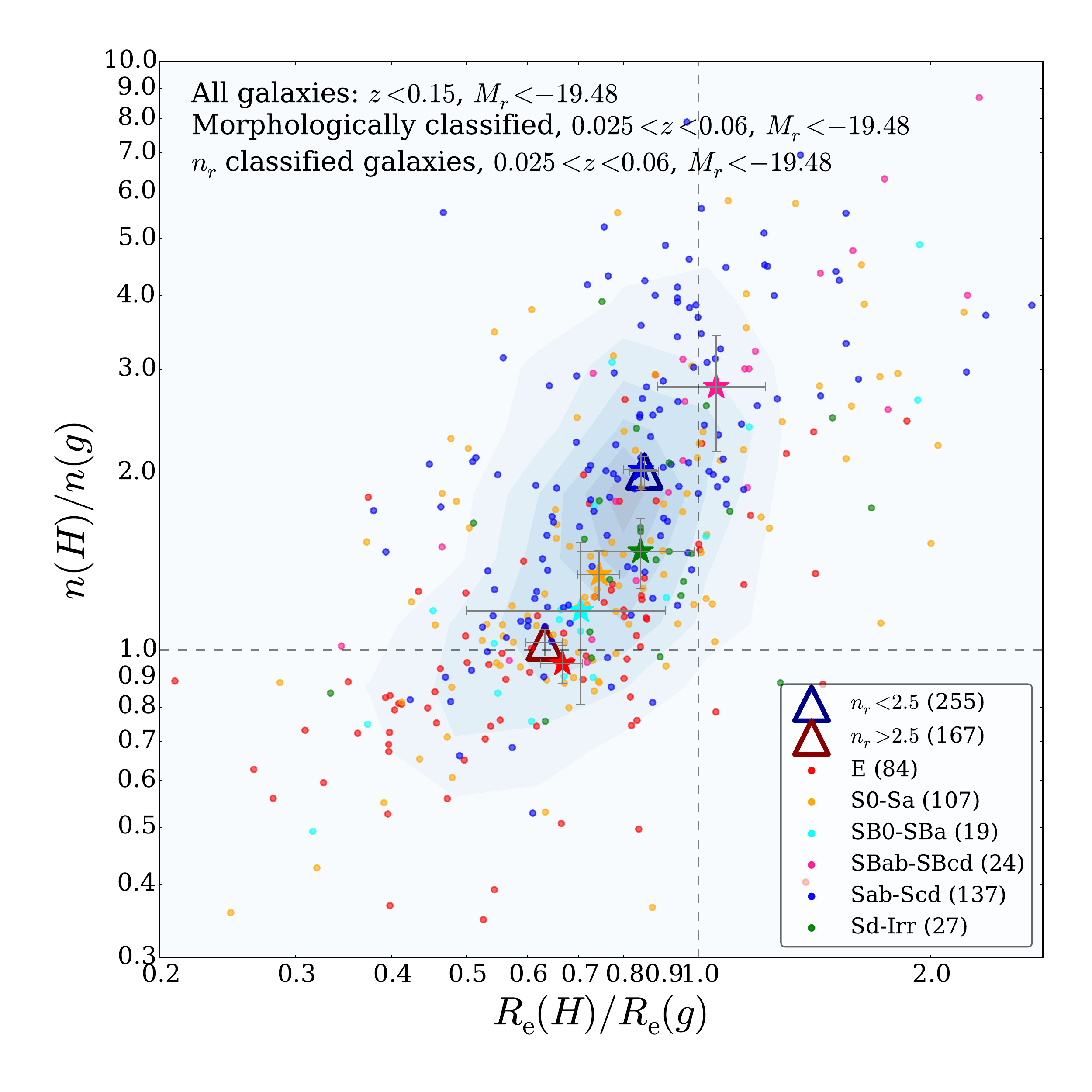}
	\caption{\N[H][g] versus \R[H][g] for galaxies in our lower-luminosity volume-limited sample.  Points and shading are as described for Fig.~\ref{fig:Morph_Vulcani}
	\label{fig:Morph_Close}}
\end{figure}

The work of V14 assumes that one can perform a meaningful division of the galaxy population using \sersic index as a proxy for structure or morphology.  Typically $n_{r} < 2.5$ galaxies are associated with structures with prominent discs, whilst $n_{r} > 2.5$ galaxies are thought to be spheroid-dominated.
To verify that our results do not strongly depend on this assumption, we now examine the distributions of \N[H][g] and \R[H][g] for samples selected by visual morphology. In Figs.~\ref{fig:Morph_Vulcani} and \ref{fig:Morph_Close} we show the wavelength dependence of $n$ and $\re$ for galaxies separated by Hubble types from the GAMA Visual Morphology catalogue \citep{Kelvin2014}.  Note that visual morphologies are only available for galaxies with $0.025 < z < 0.06$ and $M_r < -17.4$.  Figure~\ref{fig:Morph_Vulcani} shows only galaxies \B{within the overlapping redshift and magnitude regions, i.e.  $0.025 < z < 0.06$ \& $M_{r} < -21.2$ (V14 sample) or $M_{r} < -19.48$ (low-$z$ sample).}  Also plotted (as triangles) are the median locations of the high- and low-$n$ populations, selected within the same redshift and absolute magnitude limits as the morphologically-classified points.  Background contours show the distribution of galaxies in each of our volume-limited samples for which we do not have a visual morphological classification.

Despite the severe limitation in sample size that the visual morphological classifications impose, they confirm the behaviour of our high- and low-$n$ galaxy samples. In both volume-limited samples we see that the median location of the high-$n$ galaxies lies very close to the median point of the elliptical sample, whilst the median of the low-$n$ galaxies lies close to that of the Sab-Scd galaxies.  \citet{Vika2015} see identical behaviour for a sample of very low-redshift galaxies, and show that this is robust against artificial redshifting, and investigate the use of \N and \R as morphological classifiers.

Note that the offset between the late-type/low-$n$ points at low-redshift and the contoured distribution at higher redshifts appears to confirm the \B{(small)} biases inferred in Section~\ref{subsec:Galaxy_Z}.  At lower-redshift (i.e. in better quality data) we measure slightly higher \R[H][g] and \N[H][g] for low-$n$ galaxies.  Our previous results therefore somewhat underestimate the difference between the two galaxy populations.  However, even at $z=0.3$ the contrast between their behaviour is such that our conclusions are unaffected.

The finer distinctions between the Hubble types reveal some further details in the \N[H][g]--\R[H][g] figures.  Ellipticals and Sab-cd galaxies form the opposite extrema of the distributions.  S0-Sa galaxies are intermediate between the two, but closer to the ellipticals. Perhaps surprisingly, late-type disks, Sd-Irr, occupy the same intermediate region.   This may be a result of their similar one-component natures, with higher values of \N[H][g] indicative of system with comparable bulge and disc components.  \B{Note that \R and \N may be noisier for systems that depart greatly from an elliptical \sersic, and more difficult to interpret physically.  Nevertheless, measuring these properties still gives insight into their structure and allows us to make qualitative comparisons.  So, in Fig. \ref{fig:Morph_Close} we see that, on average, Sd-Irr galaxies display properties similar to late-spirals, but with less peakiness at longer wavelengths.}  It is also worth noting that the presence of a bar does not appear to greatly alter the measured wavelength dependence of galaxy structure.

\subsection{The effect of dust}
\label{subsec:dust}

Late-type galaxies contain significant quantities of dust, which can strongly affect their measured properties (e.g., \citealt{Popescu2002,Pierini2004,Holwerda2005,Holwerda2009,Mollenhoff2006}).  The effects vary with the amount and distribution of dust within the galaxy, as well as the observed inclination and wavelength.

A wide range of observations, together with detailed modelling \citep{Popescu2000}, have helped to establish a typical geometry for the distribution of dust in the disks of star-forming galaxies \citep{Tuffs2004}.  When accounting for attenuation, this dust model can be parametrised by a single dominant parameter: the central face-on optical depth in the $B$-band, $\tau$, while remaining widely applicable, at least to relatively massive spiral galaxies.  It should, however, be noted that in lower-mass galaxies the dust distribution may be significantly different \citep{Holwerda2012,Hinz2007}.   The \citet{Tuffs2004} model has been used to quantify, and potentially correct for, the effect of dust on galaxy structural parameters \citep{Mollenhoff2006,Pastrav2013a,Pastrav2013b}.

\begin{figure}
	\centering
	\includegraphics[width=0.45\textwidth]{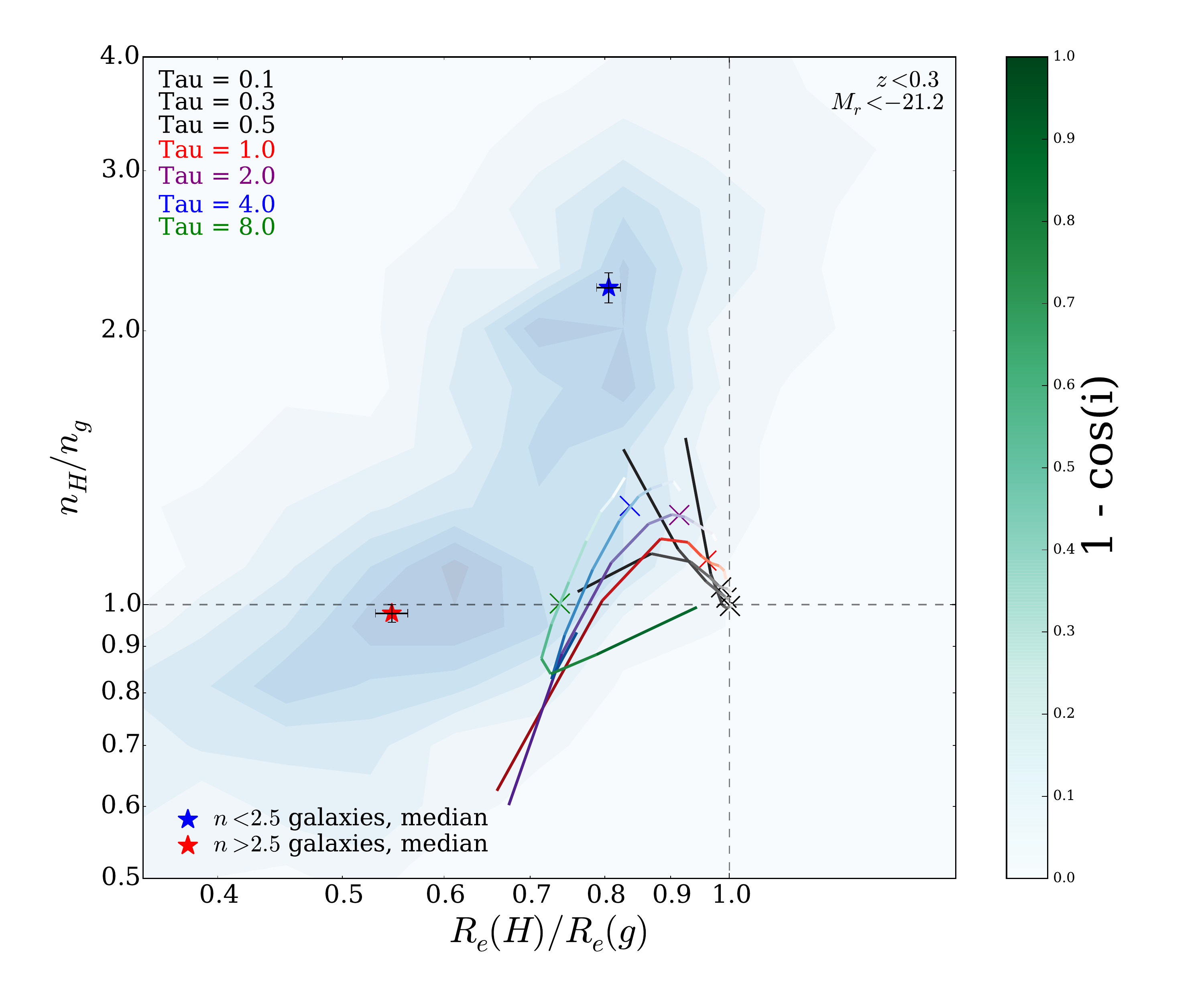}
	\caption{Dust corrected \N[H][g] vs \R[H][g] for discs. Underlayed are contours for all galaxies in the V14 sample, as in Fig. \ref{fig:Morph_Vulcani}. Overplotted are the mean locations of low- and high-$n$ galaxies within the V14 volume limited sample. Black lines show model tracks for optically thin discs having a nominal value (\N[H][g],\R[H][g])=(1,1), whilst coloured lines correspond to model tracks for optically thick discs The intensity of the colour in each line shows the inclination of the galaxy. Crosses indicate the point at which $(1-\cos{i}) = 0.5$.
	\label{fig:Dust_Vulcani}}
\end{figure}

Figure \ref{fig:Dust_Vulcani} shows the effect of dust on \N[H][g] and \R[H][g] for pure discs fit with a single-\sersic profile, as determined by \citet{Pastrav2013a}.  Their results are shown for a range of inclinations along loci corresponding to a wide variety of central dust opacities. Typical spiral galaxies have $\tau \sim 2$--$4$ \citep{Keel2001,Holwerda2005,Driver2007b,Masters2010}. Projection effects, as defined and predicted in \cite{Pastrav2013a,Pastrav2013b}, are not considered here, as their contribution is minimal compared to dust effects. As the purpose of this figure is illustrative, we have simply taken the published $K$ and $B$-band dust effects from \cite{Pastrav2013a}, without attempting the small interpolation to the $g$- and $H$-bands. These dust effects are shown as applied to a nominal `galaxy' with (\N[H][g], \R[H][g]) = (1, 1), and for each $\tau$ value we show how \N[H][g] \& \R[H][g] change with inclination. Fig.~\ref{fig:Dust_Vulcani} shows that dust always acts to decrease \R[H][g], i.e. increase apparent size more at shorter wavelengths.  It also tends to increase \N[H][g], i.e. raise the \sersic index with increasing wavelength, except for at high opacities and close to edge-on inclinations.

Note that the stellar disk emitting in the optical bands in the \citet{Tuffs2004} model actually has an intrinsic stellar population gradient, such that the scale length of the old stellar disc decreases by $\sim$30 percent from the $B$ to $K$ bands, corresponding to \R[H][g] $\sim 0.7$.  Including this intrinsic \R[H][g] would make the loci overlap with the observed distribution in Fig.~\ref{fig:Dust_Vulcani}.  One can see that for a population of pure discs with a variety of opacities and inclinations, a correlation in \N[H][g]--\R[H][g] would arise from their variation with inclination, while a scatter about that correlation would be associated with variations in opacity.  However, as one would expect, dust effects alone cannot account for the full observed \N[H][g]--\R[H][g] distribution. Late-types extend to substantially greater \N[H][g] than can be attributed to dust, presumably as a result of the presence of a central bulge that becomes more dominant at longer wavelengths.  However, there remains the possibility that more varied dust models may have more success in reproducing our observed dependence of \N and \R.  For example, the dust within lower-mass galaxies is more likely to be vertically distributed throughout the stellar disk, rather than concentrated in the central plane \citep{Holwerda2012}.  On the other hand, early-types, which generally contain little dust, must have stronger intrinsic stellar population gradients than those in the discs.

\section{Discussion}

In this paper we have studied how the wavelength dependence (from restframe $u$- to $H$- bands) of galaxy structure varies for samples selected in a variety of ways. As in V14, galaxy properties  have been measured using \megamorph techniques to fit consistent, wavelength-dependent, two-dimensional \sersic profiles in multiple wavebands simultaneously.  Our results are summarized in terms of the fractional variation in \sersic index and effective radius between the $g$- and $H$-bands, which we denote \N[H][g] and \R[H][g].

\subsection{The physical meaning of \R and \N}
Before discussing the implications of our results, and comparing to other studies, we first review the meaning of these quantities in terms of the appearance and physical structure of a galaxy.
A value around unity, for either quantity, indicates that that particular aspect of structure does not vary with wavelength.
\R gives the variation in galaxy size ($\re$) with wavelength. The vast majority of galaxies are found to have \R $< 1$, indicating that they are larger in the blue.  These galaxies therefore display redder colours in their centres, and become bluer at larger radii.  In the rarer case,  \R $> 1$, galaxies are larger in the red, and hence bluer in the centre, becoming redder with increasing radius. 

\N indicates how a galaxy's profile shape ($n$), or equivalently its central concentration, depends on wavelength. Galaxies with \N $> 1$ have a high-$n$ profile in redder bands.  They are therefore `peakier' in their centres and have more significant surface brightness at large radii. Galaxies with \N $> 1$ are thus expected to be redder in the centre and in their outskirts, and bluer at intermediate radii ($\sim \re$). In practice, the outermost red regions may be too faint to been seen in normal depth imaging, and so the visible outskirts of such a galaxy may appear blue.  However, the difference with respect to \R $< 1$ is that the colour gradient should become shallower with radius. Galaxies with \N $< 1$ obviously show the opposite trend, appearing more strongly peaked at bluer wavelengths.

Of course, both \N and \R may both depart from unity, indicating variations in both size and profile shape with wavelength. In that case, the resulting colour profile is a combination of the two behaviours. These may, to a degree, counteract (if \N and \R are correlated) or reinforce (if anti-correlated) one another.  Indeed, some of the correlation in the \N--\R plane, within subsets of the galaxy population, may be a result of this degeneracy.

We study the effect of redshift on our ability to measure \N[H][g] and \R[H][g] in Section~\ref{subsec:Galaxy_Z}.  For low-$n$ (i.e. typically late-type) galaxies we find a bias of \R[H][g] and \N[H][g] to lower values (\R[H][g] away from unity, \N[H][g] toward unity) at higher redshifts.  However, these observed biases are significantly smaller than the differences between our galaxy subsamples, particularly for low- versus high-$n$ galaxies.  The conclusions of V14 therefore are not significantly affected by the generous redshift limit adopted in that study.  The quantitative results of V14 are also relatively robust and may be compared to other studies of comparable galaxy populations in similar quality imaging. However, when comparing to much lower redshift samples of late-types, e.g. in Fig.~\ref{fig:Morph_Close}, the redshift bias must be borne in mind.

In Section~\ref{subsec:Galaxy_Mr}, the wavelength-dependence of both $\re$ and $n$ are found to depend on luminosity, with brighter galaxies generally showing stronger trends with increasing wavelength: to smaller sizes (especially for early-type galaxies) and higher-$n$ (especially late-type galaxies). 
The contrast between the behaviour of high- and low-$n$ galaxies lessens with decreasing luminosity, suggesting  greater structural similarity between (\sersic index selected) early- and late-type galaxies at lower luminosities.

\subsection{Comparison  of observed trends of R and N with other studies}
Our work is consistent with that of \citet{LaBarbera2010b}.  We find an extremely similar decrease in effective radius for red, $n > 2.5$ galaxies at the same luminosities.  Furthermore, we have shown that this is robust to redshift effects and the use of visual morphology, rather than \sersic index, to select early-types. Our results also agree with those of \citet{Kelvin2012} and \citet{Ko2005}, both of which find a nearly 40 per cent decrease in $\re$ over the corresponding wavelength range.

The existence of negative colour gradients in elliptical galaxies, particularly strong in optical-NIR colours, has been known for a long time (e.g., \citealt{Peletier1990}).  \citet{LaBarbera2010} use their \sersic models, independently fit to each wavelength, to determine colour gradients for their sample of early-type galaxies.  They find that more optically-luminous galaxies display stronger (negative) NIR-optical colour gradients. This agrees with our finding that $\re$ depends on wavelength more strongly for more luminous high-$n$ galaxies.  However, \citet{LaBarbera2010} see little variation in colour gradients with NIR luminosity or stellar mass. This is consistent with the bluer stellar population being located at larger radii, and this feature being more prominent in more luminous galaxies.

\citet{LaBarbera2012} show that, compared to optical-NIR measurements, colour gradients based only on optical bands are weaker.  They are nevertheless widely observed (e.g., \citealt{Gonzalez-Perez2011}).  Optical gradients also do not show as much variation with luminosity, being strongest for intermediate luminosities \citep{LaBarbera2012,Roche2010}. Using stacked optical (SDSS) images, \citet{DSouza2014} find strong evidence for the presence of blue (in $g-r$) halo components in various galaxy populations.  This extended component can account for around half of the light of a galaxy, being more prominent in higher-$n$ and more luminous galaxies.
Even the majority of massive early-type galaxies at $z \sim 1.5$ are found to display negative colour gradients \citep{Gargiulo2012}.

A small subset of local early-type galaxies display positive optical colour gradients, due to the presence of  blue cores associated with recent star-formation \citep{Suh2010}.  These become more prevalent at higher redshifts \citep{Ferreras2005}.  Such galaxies correspond to our high-$n$, very blue, selection.  These galaxies display $\mathcal{R}^{H}_{g} < 1$, like other early-types, implying they are more compact in the red.  However, in contrast to our other subsamples, this is combined with $\mathcal{N}^{H}_{g} < 1$, implying their profiles are more strongly peaked in the blue.  The intermediate and outer gradients of these galaxies are therefore apparently like other ellipticals, but they must display a blue excess in their cores.

Typical spirals (Sab-Scd) show less $\re$ variation with wavelength, compared to early-types, but it is still significant: a $\sim 20$ per cent decrease from $g$ to $H$.  This value has little dependence on galaxy luminosity, at least over the range we probe here.  However, note that the wavelength variation of $\re$ for late-types is somewhat susceptible to being overestimated in lower-quality data. Both early-discs (S0-Sa) and very late-types (Sd-Irr) display behaviour intermediate between ellipticals and spirals.

In terms of \sersic index, we find that early-type (high-$n$) galaxies show little variation in $n$, while  late-types substantially increase in $n$ with increasing wavelength. Again, early-discs (S0-Sa) and very late-types (Sd-Irr) fall between the two behaviours.  Our results for disc galaxies agree well with those of \citet{Kelvin2012}, who find that \sersic indices almost double from $u$ to $K$, although most of the change occurs over the optical range.

\begin{table*}
\centering

	\begin{tabular}{| l | l  l | l  l | p{10.5cm} |}
		\hline
		 & \multicolumn{2}{ c|}{`Early type'} & \multicolumn{2}{ c|}{`Late type'} & Notes \\ \hline
         & $n$ & $R_{e}$ & $n$ & $R_{e}$ &  \\ \hline
		LB10  &  1*$\uparrow$ & 28 $\downarrow$ & - & - & 5080 bright spheroids from SPIDER, $g$- to $K$-band. \citep{LaBarbera2010} \\ \hline
		Kel12 & 30 $\uparrow$ &  38 $\downarrow$ & 52 $\uparrow$ &  25 $\downarrow$ & 167,600 galaxies from GAMA, $g$- to $K$-band. \citep{Kelvin2012} \\ \hline
		V14  & 1*$\uparrow$ &  45 $\downarrow$ & 38*$\uparrow$ & 25 $\downarrow$ & 14,274 galaxies from GAMA-II G09 region, $M_{r} < -21.2, z < 0.3$, $u$- to $H$-band. \\ \hline
        L15: $10^{9}M_{\odot}$ & - &  - & - & 16 $\downarrow$ & 8399 galaxies from GAMA-II, $0.01<z<0.1$, $g$- to $K_{s}$-band. \citep{Lange2015} \\ \hline
        L15: $10^{10}M_{\odot}$ & - &  13 $\downarrow$ & - & 13 $\downarrow$ & As above. \\ \hline
        L15: $10^{11}M_{\odot}$ & - &  11 $\downarrow$  & - & - & As above. \\ \hline \hline
        K15: $M_{r} \sim -20$  & 12 $\downarrow$ &  23 $\downarrow$ & 29 $\uparrow$ & 13 $\downarrow$ & 5331 galaxies from GAMA-II G09 region, $z < 0.15$, $g$- to $H$-band (this study). \\ \hline
        K15: $M_{r} \sim -21$  & 5 $\uparrow$ & 25 $\downarrow$ & 40 $\uparrow$ & 15 $\downarrow$ & As above \\ \hline
        K15: $M_{r} \sim -22$  & 5 $\uparrow$ & 33 $\downarrow$ & 55 $\uparrow$ & 12 $\downarrow$ & As above \\ \hline
        
	\end{tabular}
	\caption{Comparison of results from previous studies.  Each number shows the percentage change in $n$ and $\re$ between the wavebands given under `notes'. $\uparrow$ = increase; $\downarrow$ = decrease; * = value has been calculated from information given in the paper; - = no data available...
	\label{table:results_comparison}}
\end{table*}

Various studies have found the \sersic indices of early-type galaxies to be generally unchanged with wavelength \citep{Kelvin2012,LaBarbera2010,LaBarbera2012}, perhaps with a slight increase in $n$ with wavelength.  The estimated changes in $n$ and $\re$ with wavelength for these, and other, studies are displayed in Table \ref{table:results_comparison}.  We do not generally see such an increase, rather a small decrease is more common in our early-type samples.  However, redshift appears to have an effect, as does luminosity.  In our lowest-redshift, brightest samples we find \sersic indices that are constant, or slightly rising, with wavelength.  Our visually-classified ellipticals also have \N[H][g] very close to unity.  On the other hand, fainter and bluer high-$n$ galaxies tend to display \sersic indices that decrease with wavelength, by several tens of per cent. \citet{LaBarbera2012} find that larger (and hence typically more luminous) early-type galaxies show stronger trends of $n$ increasing wavelength, matching the form of behaviour we see.

The redshift biases and luminosity trends we have found allow us to, at least partly, reconcile the dramatic wavelength-dependence reported by V14 with the more modest behaviour seen by some other recent studies.
\citet{Kelvin2012} find slightly weaker trends than V14, but they use a magnitude-limited sample, dominated by somewhat less luminous galaxies.  When we adopt the same selection, our results are in excellent agreement.

\citet{Lange2015} find a substantially weaker dependence of $\re$ on wavelength, though it is still significant.  We have shown that dividing galaxies by visual morphology (as in \citet{Kelvin2014}) confirms the behaviour inferred from selections based on colour and \sersic index, but severely limits the useable sample size. 

For late-type galaxies, our results are actually in good agreement with \citet{Lange2015}.  If we consider our low redshift sample, we find a similar $\sim 15$ per cent decrease in size.  At higher redshifts we appear to slightly overestimate the strength of the size decrease.
However, for early-types, there is a larger discrepancy, although the qualitative behaviour is the same.  Redshift (i.e. data-quality) appears to have little effect on our results.  Considering our $M_r \sim -20$ ($\log{M_\star} \sim 10$) selection we still find a $\sim 25$ per cent decrease in the effective radii of red high-$n$ galaxies.  This is not reduced if we limit our analysis to visually-classified ellipticals.  The wavelength-dependence results featured by \citet{Lange2015} are based on linear fits to the $\re$--$M_\star$ relation, and hence galaxies with relatively low masses, $\log{M_\star} \sim 10$, exert a strong influence.  These different analyses may have something to do with our disagreement.  However, it may also be that the difference in depth between the UKIDSS LAS and VIKING VISTA NIR imaging is responsible, as suggested by \citet{Lange2015}.  Many, but not all, of the studies that also find strong trends for early-type galaxies, as mentioned above, are also based on SDSS and UKIDSS LAS data and hence may be similarly affected. In the near future we will have the results of our methodology applied to the same SDSS plus VIKING data set, so will be better placed to evaluate this issue.

\subsection{Connection to stellar population gradients}
The wavelength dependence of structure that we have measured, i.e. \N and \R, must be driven by radial variations in stellar populations and/or dust extinction.  For early-type galaxies, many studies have found that colour gradients are caused by negative metallicity gradients.  Age is generally fairly constant with radius, or even slightly increasing, and hence acting against the metallicity trend.  The outskirts of massive early-type galaxies (ETGs) are therefore typically more metal poor and older than their cores, with weaker gradients for lower mass ETGs. (e.g., \citealt{LaBarbera2012}, and references therein).

Metallicity gradients are expected in models of monolithic collapse \citep{Worthey1995}, and this mechanism may be responsible for the gradients in the centres of today's galaxies.
However, most massive galaxies are expected to have experienced a major merger since their formation, which will have partly erased these initial gradients.
It has been discovered that early-type galaxies must at least double in effective radius between $z \sim 2$ and today (e.g., see \citealt{vanderWel2014}).
Although various processes have been proposed, the dominant mechanism appears to be minor mergers \citep{Hopkins2009,Lackner2012}.
The stellar material accreted in such mergers tends to contribute most significantly to the outskirts of the more massive galaxy, building up an outer envelope \citep{Huang2013}.  The effect is to increase effective radius over time and produce stellar population gradients.

Our measured increase in effective radius toward shorter wavelengths, which is stronger for more luminous systems, ($\mathcal{R}^{H}_{g} \sim 0.55$--$0.75$ from $M_r \sim -22$--$-20$), is consistent with observations of negative colour gradients and extended blue components \citep{LaBarbera2012}. The lack of variation in \sersic index with wavelength ($\mathcal{N} \sim 1$) suggests that the extended profile of the stellar populations seen in the blue is structurally consistent to the profile traced by those stellar populations dominating at red wavelengths, possibly indicative of a similar origin.
Together, these observations are consistent with a gradual build-up of  massive early-type galaxies through accretion of low-mass galaxies.

Lower-luminosity, and bluer, early-type (high-$n$) galaxies tend to show slightly peakier profiles in the blue, which may be an indication of an additional blue stellar component in their centres, possibly as a result of recent central star-formation.

Dust in early-type galaxies is generally not expected to be present in sufficient quantities to have a significant effect on their observed structure.  However, \citet{Rowlands2012} do find that $\sim 5$ per cent of luminous early-type galaxies are detected in the infrared, implying that they must contain significant fractions of dust (also see \citealt{Bendo2006}). It is uncertain how this dust impacts upon their measured structural parameters, but it may be responsible for some of the scatter we see in \N and \R for our early-type samples.

In late-type galaxies, dust is likely to play a more dominant role.  The apparent differences in optical structure in disc galaxies across the Hubble sequence is largely determined by the presence of young stars and dust.  The old stellar discs of spiral galaxies, as seen in the NIR, are very similar at all Hubble types \citep{Block1999}.

Modern studies of the opacity of spiral discs find that they suffer from substantial face-on and inclination-dependent extinction (e.g., \citealt{Holwerda2005,Masters2010}). Testing the model of \citet{Tuffs2004} with a large galaxy sample, \citet{Driver2007b} find good agreement and infer that bulges and the central regions of discs suffer from substantial attenuation of their optical light.  This has a significant impact on measurements of their structural parameters \citep{Mollenhoff2006,Pastrav2013a,Pastrav2013b}.  However, as we have shown in Section~\ref{subsec:dust}, dust alone cannot explain the observed wavelength trends.  It seems obvious that the two-component, bulge-disk, nature of late-type galaxies, with their contrasting stellar populations and structures, must have at least some impact on the wavelength-dependence of their overall structure.  We will investigate this further in a forthcoming paper.

\section{Summary}

The variation of galaxy structure with wavelength reveals the connections between the stellar populations within a galaxy and their spatial distributions.  Specifically, we consider the change in \sersic index and effective radius from $g$- to $H$-band, which we denote \N and \R.  As shown by V14, the majority of early-type (high-$n$) galaxies show little variation in their \sersic index with wavelength ($\mathcal{N} \sim 1$), but are significantly smaller at longer wavelengths ($\mathcal{R} < 1$).  This behaviour is suggestive of a structure formed by similar (violent) mechanisms being apparent at all wavelengths, but with a scale that strongly depends on the colour of the stellar population.  On the other hand, late-type galaxies (low-$n$) display a substantial increase in \sersic index with wavelength ($\mathcal{N} > 1$), suggesting a variation in the type of structure dominating the light at different wavelengths, but less variation in size ($\mathcal{R} \la 1$).  Very blue, high-$n$, galaxies present an interesting contrast, having significantly higher \sersic indices in the blue, perhaps indicating the presence of a peaky blue component.

In this paper we have followed V14, using optical-NIR imaging and redshifts from the GAMA survey \citep{Driver2009}, with multi-band single-\sersic fits performed using tools developed by the MegaMorph project \citep{Bamford2012}, in order to study four further aspects of the wavelength-dependence of galaxy structure.  We consider the distributions of \N and \R for a variety of volume-limited samples subdivided by $u-r$ colour, \sersic index and morphology. The main conclusions from our analysis are:

\begin{itemize}
	\item We have verified that our measurements of \N and \R are robust to the effects of redshift, strengthening our earlier results (V14).  Early- and  and late-type galaxies (selected using \sersic index) present strongly contrasting behaviour in terms of the wavelength-dependence of their structure. Out to $z \sim 0.3$ we see no substantial changes in \R or \N, particularly when compared to the striking differences between galaxy subsamples.  What small biases are present, suggest that at higher redshift (lower S/N and poorer resolution) we may be slightly overestimating the offset of \R from unity, and underestimating the offset of \N from unity, for low-$n$ galaxies.  Both of these effects act to reduce the apparent difference between the high- and low-$n$ populations.  The estimated contrast is therefore even more pronounced than that determined by V14.
    \item The strengths of \N and \R depend on galaxy luminosity:
    
    \begin{itemize}
    \item High-$n$ galaxies with lower luminosities have \R closer to unity than their brighter counterparts. Most high-$n$ galaxies show a peak in \N around unity for all luminosities, supporting the suggestion of V14 that these are single-component objects.
    \item Low-$n$ galaxies display weaker, mixed trends in \R than high-$n$ galaxies. There is a striking trend in the \N distributions of low-$n$ galaxies with luminosity: brighter objects display \N further from unity, for all colour subsamples.
    \end{itemize}
    
    \item The interpretations of V14 are supported by visual morphological classifications of a low-redshift subset of our sample from \citet{Kelvin2014}; our low-$n$ samples share the same part of the \N--\R diagram as Sab/Scd galaxies, whilst the high-$n$ samples follow the distribution of elliptical galaxies.  Both early- and late- disc galaxies occupy intermediate regions of the \N--\R plane, confirming that the extreme values for intermediate types are related to their two-component nature.
    \item Results from fitting dusty galaxy models \citep{Pastrav2013a} indicate that some of the wavelength-dependence of disc galaxy structure may be attributable to dust. The natural distribution of disc inclinations may account for the trend seen in the \N--\R plane, while varying central dust opacities may account for the scatter in this relation.  However, pure dusty discs cannot reach the values of \N observed for intermediate-type spirals.
\end{itemize}

Further improving our understanding of the wavelength-dependence of galaxy structure will require more detailed consideration of the two-component nature of galaxies.  We are currently exploring this topic using bulge-disc decompositions, and these results will be the subject of Kennedy et al. (in prep.).

\section{Acknowledgements}

This paper is based on work made possible by NPRP award 08-643-1-112 from the Qatar National Research Fund (a member of The Qatar Foundation).

GAMA is a joint European-Australasian project based around a spectroscopic campaign using the Anglo-Australian Telescope. The GAMA input catalogue is based on data taken from the Sloan Digital Sky Survey and the UKIRT Infrared Deep Sky Survey. Complementary imaging of the GAMA regions is being obtained by a number of independent survey programs including GALEX MIS, VST KiDS, VISTA VIKING, WISE, Herschel-ATLAS, GMRT and ASKAP providing UV to radio coverage. GAMA is funded by the STFC (UK), the ARC (Australia), the AAO, and participating institutions. The GAMA website is http://www.gama-survey.org/.

SPB gratefully acknowledges the receipt of an STFC Advanced Fellowship.

BV acknowledges the financial support from the World Premier International Research Center Initiative (WPI), MEXT, Japan and  the Kakenhi Grant-in-Aid for Young Scientists (B)(26870140) from the Japan Society for the Promotion of Science (JSPS)

\B{We thank the anonymous referee for their comments, which helped to improve the presentation of this paper.}

\bibliographystyle{mn2e}
\bibliography{4Dec14}

\end{document}